\documentclass[pra,twocolumn,superscriptaddress,showpacs,floatfix]{revtex4-2}

\usepackage{graphics}
\usepackage{graphicx}
\usepackage{bm}
\usepackage{amsmath}
\usepackage{amsfonts}
\usepackage{amssymb}
\usepackage{latexsym}
\usepackage{color}

\begin{document}
\title{Quantum physics cannot be captured by classical linear hidden variable theories \\ even in the absence of entanglement}
\author{Kawthar Al Rasbi}
\affiliation{The School of Physics and Astronomy, University of Leeds, Leeds LS2 9JT, United Kingdom}
\affiliation{The Department of Physics, Sultan Qaboos University, Sultanate of Oman}
\author{Lewis A.~Clark}
\affiliation{Centre for Quantum Optical Technologies, Centre of New Technologies, University of Warsaw, Banacha 2c, 02-097 Warsaw, Poland}
\author{Almut Beige}
\affiliation{The School of Physics and Astronomy, University of Leeds, Leeds LS2 9JT, United Kingdom}

\date{\today}

\begin{abstract}
Recent experimental tests of Bell inequalities confirm that {\em entangled} quantum systems cannot be described by local classical theories but still do not answer the question whether or not quantum systems could in principle be modelled by linear hidden variable theories. In this paper, we study the quantum trajectories of a single qubit that experiences a sequence of repeated generalised measurements. It is shown that this system, which constitutes a Hidden Quantum Markov Model, is more likely to produce complex time correlations than any classical Hidden Markov Model with two output symbols. From this, we conclude that quantum physics cannot be replaced by linear hidden variable theories. Indeed, it has already been recognised that not only entanglement but also non-classical time correlations of quantum systems with quantum feedback are a valuable resource for quantum technology applications.
\end{abstract}

\maketitle 

\section{Introduction} \label{sec1}

Entanglement, as defined by Erwin Schr\"odinger \cite{schroedinger}, is considered by many the ``essence of quantum physics'' \cite{brukner2021essence} and the origin of ``spooky action at a distance" \cite{Gudder}, and therefore receives a lot of attention, especially in recent research into quantum information processing. Many quantum technology applications, from quantum cryptography to quantum metrology and quantum computing, require entanglement as a resource. It is therefore not surprising that entanglement is also often at the centre of a debate which tries to draw a clear line between quantum and classical physics. Already in 1935, scientists like Einstein, Podolsky and Rosen asked the questions whether or not the dynamics of quantum systems could be described by classical hidden variable theories \cite{EPR}. They were hoping that quantum physics was simply a way of dealing with a lack of knowledge rather than indicating the need for an alternative, non-deterministic approach to physics.

In 1964, Bell constructed an inequality that could be violated by entangled quantum systems, but not by performing measurements on two individual classical particles with local hidden properties \cite{bell1964einstein}. Suddenly, the question of physics being either quantum or classical was no longer just a matter of interpretation. It was now possible to verify and quantitatively measure the strangeness of quantum systems. Over the years, several tests of Bell inequalities \cite{CHSH} have been performed \cite{Clauser,Aspect,zeilinger,gisin} and a strong case has been made for the reality of quantum physics and the existence of entanglement. Eventually, in 2015, additional loopholes of previous Bell tests were closed \cite{1,2,3} and quantum physics became widely accepted not only as a highly efficient but also necessary approach. Although it is still possible to describe quantum systems by linear hidden variable theories, it was concluded that such classical theories would have to be at least non-local.

However, entanglement is not the only strange property of quantum systems. Another characteristics that quantum systems do not share with classical systems is that their measurement outcomes can be discrete even when their internal states are continuous. For example, weak light arriving at a detector either causes a click or no click. Hence performing a measurement on a quantum systems must alter its state, thereby causing a quantum jump to occur \cite{Bohr}. Without quantum jumps, repeating a measurement on a quantum system might not yield the same outcome, which would mean that the outcome of the previous measurement was meaningless.  This is in contrast to classical physics where a measurement reveals information but does not cause a physical system to change.

In 1975 Dehmelt pointed out that driving a single three-level atom with appropriate laser fields can lead to macroscopic quantum jumps \cite{Dehmelt}. These are a random sequence of long periods of constant fluorescence interrupted by long periods of no fluorescence. The light and dark periods of a blinking atom occur on macroscopic time scales and are a manifestation of very persistent time correlations. Subsequently, the existence of macroscopic quantum jumps has been experimentally verified by several groups \cite{4,5,6} and theoretical models were developed to accurately predict the statistical properties of their trajectories \cite{7,8,9,10}. Since there is a large variety of classical stochastic processes, it is hard to argue that the time correlations in these quantum experiments, were non-classical. Quantum jump experiments were mainly seen as an interesting way of illustrating the stochastic nature of quantum physics \cite{Blatt}. Eventually, it was pointed out that macroscopic light and dark periods could be used to herald the generation of entangled atomic states \cite{Metz,Metz2}. 

To capture the non-classicality of time correlations, temporal Bell inequalities \cite{LG,PM,BV,Budroni,Zych,DV,milz2021genuine} were introduced, which could only be violated by quantum but not by classical stochastic processes. For example, in 2004, Brukner {\em et al.}~\cite{BV} coined the term ``entanglement in time." To verify the non-classicality of their measurement correlations, authors looked for examples of causality violations, since classical systems are always causal \cite{OO,Goswami}. For example, quantum switches with applications in quantum communication require a single quantum system to simultaneously experience {\em two or more quantum channels} such that the order of cause and effect becomes obscured \cite{switch}. Indeed, it has already been shown that quantum switches and other systems which can violate causality provide interesting additional resources for quantum technology applications, like quantum computing and quantum communication \cite{QT1,QT2,QT3,QT4,QT5}.

The purpose of this paper is to emphasise that temporal quantum correlations can occur even when a quantum system as simple as a single qubit experiences only {\em a single quantum channel}. As we shall see below, sequential generalised measurements can generate non-classical time correlations, even in the absence of causality violations and entanglement. From this, we conclude that stochastic quantum processes cannot be captured by classical linear hidden variable theories, thereby imposing a much tighter boundary on what behaviour can be considered {\em quantum}. In good agreements with Refs.~\cite{korzekwa2021quantum,elliott2018,blank2021quantum,elliott2020,elliott2022,vieira2022temporal,garner2017provably,milz2021quantum}, we find that quantum physics can reduce the complexity and memory needed, for example, when simulating certain stochastic processes. 

As in Ref.~\cite{vieira2022temporal}, we consider in the following a single qubit which experiences the same generalised measurement many times. Its measurement outcomes ``$A$" and ``$B$" are recorded and we then study the stochastic properties of the generated random sequences. As we shall see below, such a system can be classified as a Hidden Quantum Markov Model (HQMMs) with one memory qubit. HQMMs \cite{Wiesner,monras2010hidden,clark2014hidden,marzen2015informational,cholewa2017quantum} are quantum versions of stochastic generators which became known as Markov Models (MMs) \cite{norris1998markov} and as Hidden Markov Models (HMMs) \cite{first,rabiner1986introduction,rabiner1989tutorial,Eddy1996,dymarski2011hidden}. A clear advantage here is the ability of quantum systems to maintain correlations over longer periods of time than classical systems, thus enabling them to exhibit more exotic behaviour even with a comparable amount of resources. Because of this, HQMMs already found applications in quantum machine learning \cite{srinivasan2018learning,markov2022implementation} and in simulating open quantum systems \cite{wood2011tensor,binder2018practical}. Moreover, HQMMs based on single-mode coherent states can lead to a violation of the standard quantum limit and a quantum advantage in quantum metrology applications even in the absence of entanglement \cite{QJM,Clark2019,QJM3}.

In the following, we study the stochastic properties of the output sequences of a relatively large class of HQMMs with a single memory qubit and compare them with the stochastic properties of the output sequences of MMs and HMMs with two output symbols ``$A$" and ``$B$". Despite looking only at a subset of all possible HQMMs, we find that HQMMs can generate stochastic processes with stronger time correlations than all possible MMs and HMMs with two output symbols, even when we allow for HMMs with hidden memories of any size. More concretely, we find that increasing the memory of a HMM does not increase its complexity. In MMs and HMMs, the presence of temporal correlations in consecutive outputs cannot be sustained for very long and previous output symbols are in general much more quickly forgotten than in the case of HQMMs.

This paper contains five sections. In Section \ref{sec2}, we introduce and define MMs, HMMs and HQMMs, and introduce the notation which is used throughout the paper. Afterwards, in Section \ref{sec3}, we parametrise all three machines and derive analytical expressions for output probabilities of certain stochastic sequences. Due to the ergodicity of the considered stochastic generators, we assume that they possess a stationary state and only consider word probabilities for the case when the machine transitioned into its stationary state. After introducing all the necessary theoretical characterisation of the three models, we present a numerical comparison of their complexity in section \ref{sec4}.  Finally, we summarise our findings in Section \ref{sec5}.

\section{Hidden Quantum Markov Models and their classical counterparts} \label{sec2}

\begin{figure}[t]
\centering
\includegraphics[width = 0.40 \textwidth]{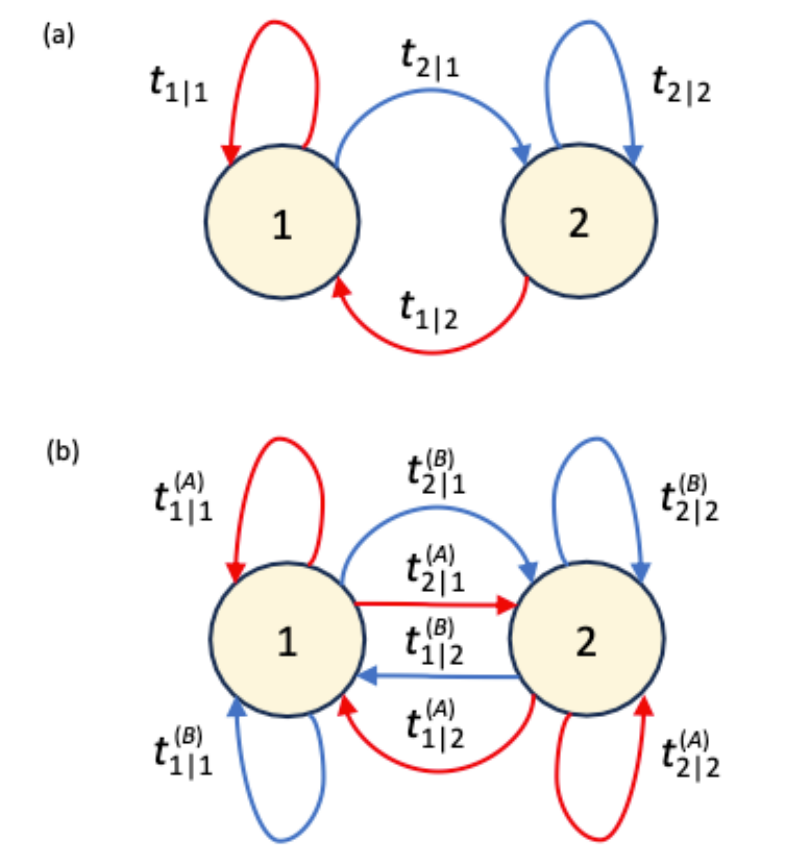}
\caption{
(a) Schematic view of a Markov Model with two states $1$ and $2$. Here $t_{j|i}$ denotes the probability of the machine to transition from state $i$ into state $j$. During each transition, an output symbol is created. All red arrows are accompanied by the generation of an ``$A$", while the blue arrows correspond to transitions which generate a ``$B$". In this way, the obtained output symbol is a clear indication of the state of the machine. The state of the machine is therefore not hidden.
(b) Schematic view of a Markov Model with two states $1$ and $2$. Here $t_{j|i}$ denotes the probability of the machine to transition from state $i$ into state $j$. During each transition, an output symbol is created. All red arrows are accompanied by the generation of an ``$A$", while the blue arrows correspond to transitions which generate a ``$B$". In this way, the obtained output symbol is a clear indication of the state of the machine. The state of the machine is therefore not hidden.}
\label{fig1}
\end{figure}

Markov Models (MMs), which are also known as Markov chains, are memoryless generators of stochastic processes. One way of simulating more complex stochastic processes is to replace these machines by Hidden Markov Models (HMMs). Alternatively, more complex stochastic sequences can be generated by Hidden Quantum Markov Models (HQMMs) which take advantage of quantum physics. In this subsection, we have a closer look at the definitions of all three machines. For simplicity, we restrict ourselves to machines with only two possible outputs, $A$ and $B$, but set the notation out that can be generalised to higher orders. 

\subsection{Markov Models} \label{sec2.3}

Discrete-time MMs evolve on a coarse grained time-scale and each time step $\Delta t$ of their dynamics is dominated by a random transition from a state $i$ to another state $j$. At the end of each time step, a measurement is performed and an output symbol is generated which indicates the current state of the machine \cite{norris1998markov}. The dynamics of discrete-time MMs and their measurement outputs depend solely on the previous state $i$ but not on the history of the machine, which is known as its \emph{Markov property}. Discrete MMs are therefore fully characterised by 3-tuples $({\cal S},{\cal T},{\boldsymbol p}(0))$ where ${\cal S}$ describes the available state space, ${\cal T}$ specifies transition probabilities and the vector ${\boldsymbol p}(0)$ contains all the probabilities to find the machine in a certain initial state. Here we only consider stochastic machines with two possible outputs $A$ and $B$ and two internal states $1$ and $2$. As illustrated in Fig.~\ref{fig1}(a), preparing the machine in $1$ or in $2$ generates the output $A$ or $B$, respectively. Suppose $S(n) = i$ is the state of an individual machine after $n$ time steps. Then
\begin{eqnarray} \label{MC}
S(n+1) &=& j
\end{eqnarray}
with transition probability $t_{j|i}$. The current state $S(n+1)$ does not contain any information about the previous nor the initial state of the machine. 

However, if measurement outcomes are ignored or if ensemble averages for a large number of MMs are considered, the states $S(n)$ of individual MMs remain unknown. In this case, we describe the state of the machine by a two-dimensional vector ${\boldsymbol p}(n)$ of the form
\begin{eqnarray} \label{statevec}
{\boldsymbol p}(n) = \left( \begin{array}{c}  p_1 (n) \\ p_2 (n) \end{array} \right) \, .
\end{eqnarray}
Here $p_i(n)$ is the probability of finding the machine after $n$ time steps in state $i$ and $p_1 (n) + p_2 (n) = 1$. Ignoring output symbols, the dynamics of the state vector ${\boldsymbol p}(n)$ can be described by a single transition matrix $T \in {\cal T}$ which is defined such that
\begin{eqnarray}
{\boldsymbol p}(n+1) =T \, {\boldsymbol p}(n) \, .
\label{MC2}
\end{eqnarray}
The operator $T$ contains all the probabilities $ t_{j|i}$ that govern the dynamics of the MM. For example, for the two-state Markov chain in Fig.~\ref{fig1}(a), we have 
\begin{eqnarray} \label{TMM}
T= \left( \begin{array}{cc}  t_{1|1} & t_{1|2} \\ t_{2|1} & t_{2|2} \end{array} \right) \, .
\end{eqnarray}
As we shall see in Section \ref{sec4}, the relative simplicity of MMs has the drawback of resulting in only weak correlations in their output sequences.

\subsection{Hidden Markov Models} \label{sec2.2}

Different from MMs, HMMs are stochastic generators whose output symbols do {\em not} reveal their state. The internal state remains {\em hidden}. Nevertheless, HMMs too obey the Markov property and output symbols and transitions depend only on the current state of the machine. Hence HMMs are characterised by 4-tuples $({\cal S},{\cal T},{\cal O},{\boldsymbol p}(0))$. Again ${\cal S}$ denotes the relevant state space and ${\boldsymbol p}(0)$ describes the initial state. However ${\cal T}$ now contains probabilities for transitioning from a certain state $i$ into a certain state $j$ while creating a certain output symbol. Moreover, ${\cal O}$ specifies the possible measurement outcomes. Notice that there are two types of HMMs which are called Moore and Mealy \cite{clark2014hidden}. In the first case, the generated stochastic output depends only on the current state of the machine. In the latter case, it depends on the current state and on the observed output symbol, as illustrated in Fig.~\ref{fig1}(b). In this paper, we focus on Mealy HMMs  which contain Moore HMMs as a subset. The reason for looking at these is that we are interested in comparing quantum machines to the most general possible classical stochastic generators. 

Here we are especially interested in machines with only two possible outputs, namely $A$ and $B$. However, in the case of HMMs, this does not restrict the number of internal states $N$. In the following, we denote the hidden states of the HMM by $i$, with $i$ varying from $1$ and $N$. Suppose $t^{(m)}_{j|i}$ is the probability for a HMM prepared in state $i$ to transition to $j$ while generating output $m$. In this case, given $S(n)=i$, we again find that $S(n+1) = j$, as in Eq.~(\ref{MC}). The probability to obtain this state now equals the sum of the probabilities of generating an $A$ and a $B$,
\begin{eqnarray} \label{tij}
t_{j|i} = t^{(A)}_{j|i} + t^{(B)}_{j|i} \, .
\end{eqnarray}
Since an external observer has access to the outputs of the machine but not to its hidden states, we now have two transition matrices $T_m$ with $m=A,B$ such that
\begin{eqnarray}
T_m = \left( \begin{array}{cccc} 
t^{(m)}_{1|1} & t^{(m)}_{1|2} & \ldots &  t^{(m)}_{1|N} \\
t^{(m)}_{2|1} & t^{(m)}_{2|2} & \ldots &  t^{(m)}_{2|N} \\
\vdots & \vdots & & \vdots \\
t^{(m)}_{N|1} & t^{(m)}_{N|2} & \ldots &  t^{(m)}_{N|N} \\
\end{array} \right) \, .
\end{eqnarray} 
Using this notation, the state vector ${\boldsymbol p}(n+1)$ with coordinates $p_i(n+1)$ which represent the probability of the machine being in state $i$ after $n+1$ steps equals
\begin{eqnarray} \label{state}
{\boldsymbol p}(n+1) &=& T_m \, {\boldsymbol p}(n) / {\rm Pr}_{n} (m) \, ,
\end{eqnarray}
if the state of the machine equalled ${\boldsymbol p}(n)$ after $n$ steps and output $m$ was obtained in step $n+1$. Here ${\rm Pr}_{n} (m) $ denotes the probability of obtaining the output $m$ in step $n$. Different from MMs, the state vectors ${\boldsymbol p}(n)$ are now real vectors of dimension $N$.

For example, we can now calculate the probability of a HMM being in a certain state $i$ after $n$ steps when all measurement outputs are ignored. As in the previous subsection, in this case, the dynamics of the state vectors of the HMM can be described by a transition matrix $T$, 
\begin{eqnarray} \label{TTT}
T &=& T_A + T_B \, ,
\end{eqnarray}
which is now the sum of the two sub-transition matrices $T_A$ and $T_B$. Using this notation
\begin{eqnarray} 
{\boldsymbol p}(n+1) &=& T \, {\boldsymbol p}(n) \, ,  
  \label{MC2x}
\end{eqnarray}
in analogy to Eq.~(\ref{MC}). The matrix elements of the total transition matrix $T$ are the $t_{j|i}$ in Eq.~(\ref{tij}). If $\eta = (1,1, \ldots ,1)$ is a row vector with all $N$ coordinates equal to 1, then the probability ${\rm Pr}_{n+1} (m) $ of getting output $m$ after $n+1$ steps equals
\begin{eqnarray} \label{prob}
{\rm Pr}_{n+1} (m) &=& \eta \, T_m \, {\boldsymbol p}(n) \, .
\end{eqnarray}
We now have all the information needed to simulate all possible individual trajectories of a given HMM with $N$ internal states.

\subsection{Hidden Quantum Markov Models} \label{sec2.1}

To obtain quantum versions of HMMs, all we need to do is to replace their hidden states $i$ by quantum states $|i \rangle$. However, being quantum, the allowed internal states of a HQMM are {\em not} discrete but continuous. In general, the hidden quantum memory is in a linear superposition of a finite number of discrete quantum states. More concretely, the state $|\psi (n) \rangle$ of a HQMM after $n$ steps can always be written as 
\begin{eqnarray} \label{oma}
|\psi (n) \rangle &=& \sum_{i=1}^N c_{i}(n) \, |i \rangle \, ,
\end{eqnarray}
where the $c_{i}(n)$ denote complex coefficients with 
\begin{eqnarray} \label{oma2}
\sum_{i=1}^N  |c_{i}(n)|^2 =1 \,. 
\end{eqnarray}
As before, the output symbol and the transition of a HQMM depends only on its current internal state and transitions between subsequent states $|\psi(n) \rangle$ are again specified by linear stochastic process transition matrices \cite{monras2010hidden}. More concretely, given output symbol $m$ is measured, the state of the HQMM changes such that
\begin{eqnarray} \label{13}
|\psi(n+1) \rangle &=& \frac{K_m |\psi(n) \rangle}{\| K_m |\psi(n) \rangle\|} \, .
 \end{eqnarray}
 The probability to obtain this outcome for the given initial state $|\psi(n) \rangle$ now equals
\begin{eqnarray}
{\rm Pr}_{n+1}(m) = \| K_m \, |\psi(n) \rangle  \|^2
\end{eqnarray}
which is different from the probability ${\rm Pr}_{n+1}(m)$ in Eq.~(\ref{prob}). The operator $K_m$ in the above equations is a so-called Kraus operators and must obey certain measurement constraints \cite{Kraus}. For example, here we are especially interested in HQMMs with two output symbols, $A$ and $B$. In this case, probabilities for the two possible measurement outcomes only add up to one when 
\begin{eqnarray} \label{constraint}
K^\dagger_A K_A + K^\dagger_B K_B &=& I 
\end{eqnarray}
where $I$ denotes the identity operator. In this case, the HQMM uses only a single qubit as its internal memory. This means, the hidden state of the machine belongs to a two-dimensional state space ($N=2$).

Again, if the measurement outputs of the machine are ignored, we cannot know the states $|\psi(n) \rangle $ of individual HQMMs, even when their initial state $|\psi(0) \rangle$ is known. In this case, the  probability vectors ${\boldsymbol p}(n)$ of HMMs need to be replaced by density matrices $\rho (n)$. These describe the quantum states of the memory averaged over a large ensemble of individual machines and allow us to predict the dynamics of expectation values, like the probability of finding the HQMM at a certain time step in a certain internal state. Instead of Eq.~(\ref{13}), the dynamics of the HQMM is now given by
\begin{eqnarray} \label{qsteps}
\rho(n+1) &=& {\cal K} (\rho(n))
\end{eqnarray}
with the superoperator ${\cal K}$ defined such that
\begin{eqnarray} \label{defK}
{\cal K} (\rho(n)) &=& \sum_{m=A,B}  K_m \rho(n) K^\dagger_{m} \, .
\end{eqnarray}
In other words, the Kraus operators $K_A$ and $K_B$ replace the transition matrices $T_A$ and $T_B$ which we introduced in the previous subsection. For example, the probability ${\rm Pr}_{n+1}(m)$ of generating an output $m$ in step $n+1$ now equals
\begin{eqnarray}
{\rm Pr}_{n+1}(m) &=& {\rm Tr} \left( K_m \rho(n) K^\dagger_{m} \right) \, ,
 \end{eqnarray}
where the trace, ${\rm Tr}$, denotes the sum of all the diagonal matrix elements. 

\section{Parametrisation, stationary states and different observable properties} \label{sec3}

\begin{figure}[t]
\includegraphics[width=0.48\textwidth]{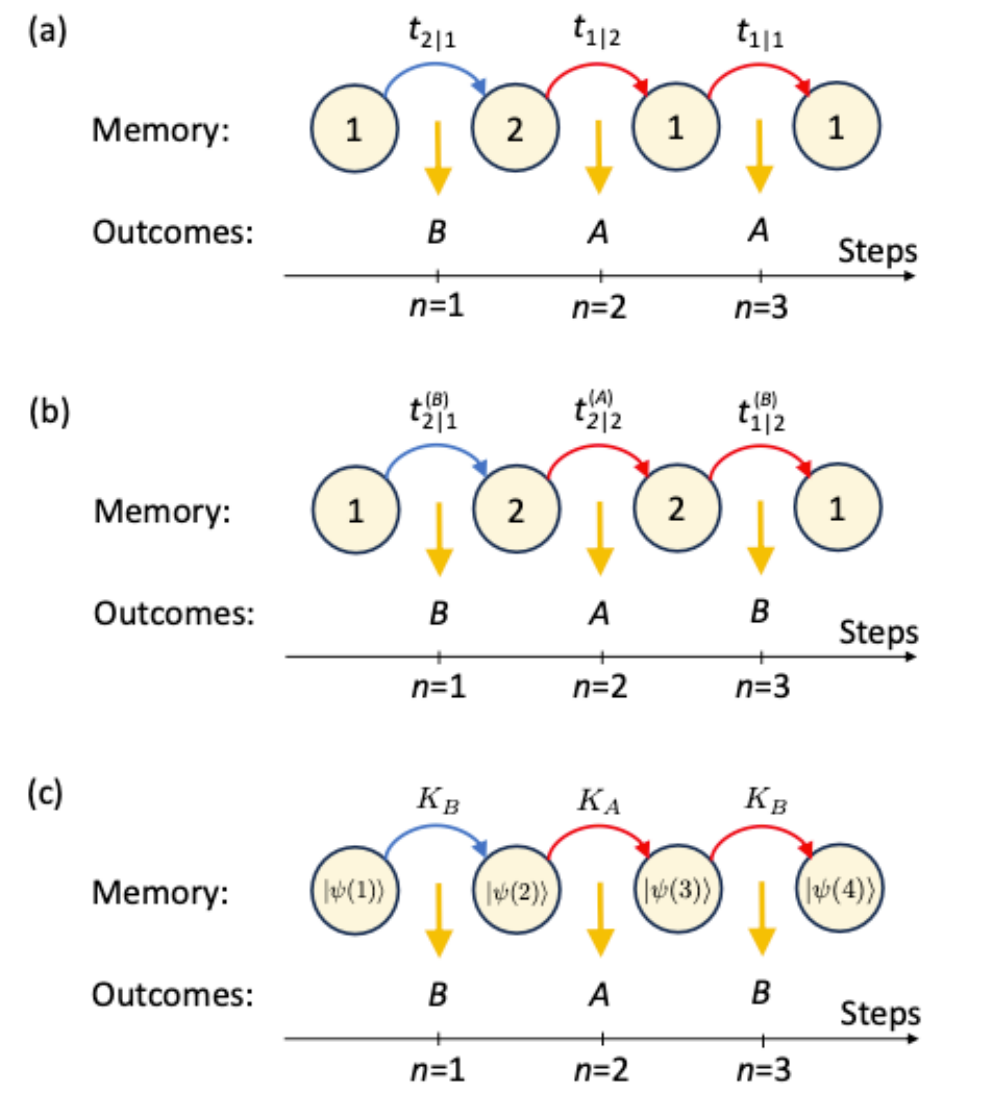} 
\caption{A comparison of the mechanisms which lead to the generation of output symbols in the cases of MCs, of HMMs and of HQMMs. 
(a) A Markov chain has two states $A$ and $B$ and the transition between states over time is governed by the transition probability matrix. The grey arrows shows all possible routes of evolving the Markov chain with time while the red arrow shows a single trajectory.
(b) Hidden Markov model evolution over time. Here the Markov chain has two states $A$ and $B$ and the transition between states over time is governed by the transition probability matrix. The Diagram shows one possible routes of evolving the Markov chain with time.
(c) Hidden quantum Markov model evolution over time. Here the hidden quantum Markov model has two states $A$ and $B$ and the transition between states over time is governed by the transition probability matrix. The diagram shows all possible routes of evolving the Hidden quantum Markov model with time.}
\label{fig2}
\end{figure}

In this section we parametrise the machines that we introduced in the previous section and highlight the constraints that must be satisfied by each model. In addition, we determine stationary state distributions whenever possible and calculate the probabilities for certain output sequences. For simplicity, we assume in the following that all three machines are ergodic due to their finite size and therefore possess a stationary state. The calculated `word probabilities' (probability of a specific output sequence) therefore apply to the outputs of large ensembles of machines, which all have already reached their stationary states. They also apply to all the words generated by a single machine when averaged over an infinitely long trajectory of output symbols, like the ones illustrated in Fig.~\ref{fig2}.

\subsection{Markov Models} \label{sec3.1}

From probability theory, we know that the matrix elements of the transition matrix $T$ in Eq.~(\ref{TMM}) are all between zero and one. In addition, they must obey the condition
\begin{eqnarray}
t_{1|i}+ t_{2|i} &=& 1  
\end{eqnarray}
for $i = 1,2$ to ensure that the machine always transitions into one of its two available states $1$ and $2$. Since $T$ has four matrix elements and there are two constraints, we can parametrise one-bit MMs using only two independent parameters $p$ and $q$. More concretely, we write $T$ in the following as
\begin{eqnarray} \label{TMM2}
T &=& \begin{pmatrix}
p & 1-q \\
1-p & q
\end{pmatrix} 
 \end{eqnarray}
with parameters $p,q \in (0,1)$. Here $t_{1|1} = p$, $t_{2|2} = q$, $t_{2|1} = 1-p$ and $t_{1|2} = 1-q$. In the next section, we will choose $p$ and $q$ randomly to sample a large set of all possible random machines and study the properties of their output sequences.

The possible output sequences of each machine depend somewhat on their respective initial state ${\boldsymbol p}(0)$. However, if the machine is ergodic and its outputs are ignored for a certain initial minimum amount of time, it soon assumes a stationary state ${\boldsymbol p}_{\rm ss} = (p_1, p_2)^{\rm T}$ with
\begin{eqnarray} \label{Tpi}
T \, {\boldsymbol p}_{\rm ss} &=& {\boldsymbol p}_{\rm ss}  \, . 
 \end{eqnarray}
Using Eq.~(\ref{TMM2}) and taking into account that $p_{1}$ and $p_{2}$ must add up to one, one can show that 
\begin {eqnarray} \label{MMss}
p_1 = \frac{1-q}{2-p-q} \, , ~~ p_2 = \frac{1-p}{2-p-q} \, .
\end {eqnarray}
These probabilities tell us how likely it is to find a certain machine which is characterise by $p$ and $q$ either in $1$ or in $2$. Hence they also equal the probabilities to obtain the outputs $A$ and $B$ at any time $n$, respectively, if the previous outputs of the machine are unknown and cannot be taken into account. 

Suppose the machine has initially been prepared in its stationary state ${\boldsymbol p}_{\rm ss}$. Then the probability $P(i_1 i_2 \ldots i_m)$ of obtaining the output sequence $i_1 i_2 \ldots i_m$ of length $m$ simply equals
\begin{eqnarray} \label{23}
P(i_1i_2 \ldots i_m) &=& t_{i_m|i_{m-1}} \ldots t_{i_3|i_2} t_{i_2|i_1} \, p_{i_1} 
\end{eqnarray}
with $p_{i_1}$ given in Eq.~(\ref{MMss}). For example, the probability of creating a word of length $m+2$ which starts and ends with the output symbol $A$ and otherwise only contains $B$'s equals
\begin{eqnarray} \label{24}
P_m (AB \ldots BA) &=& {(1-p)(1-q)^2 \over 2-p-q} \, q^{m-1} \, .
\end{eqnarray}
Alternatively, we could ask for example the question, what is the probability of a word of length $m+2$ to start and to end with the output symbol $A$. This probability equals
\begin{eqnarray} \label{26}
P_m(A* \ldots *A) &=& \frac{1-q}{2-p-q} \, (1,0) \, T^{m-1} \left( \begin{array}{c} 1 \\ 0 \end{array} \right) \, . ~~
\end{eqnarray}
Probabilities like the above ones are a measure for the complexity of the machine. For example, the probability $P_m(A *\ldots *A)$ can tell us how long correlations persist in the output sequences of a machine. For large $m$, the probability $P_m(A* \ldots * A)$ tends to $p_1$ and any knowledge about having been prepared in $1$ exactly $m+1$ steps earlier is lost.

\subsection{Hidden Markov Models} \label{sec3.2}

As we have seen in Section \ref{sec2.1}, the description of HMMs with two outputs and $N$ internal states requires $2$ transition matrices $T_m$ with $N^2$ matrix elements $t^{(m)}_{j|i}$. To identify the number of independent parameters that is needed to numerically simulate all possible HMMs, we first notice that the matrix elements $t^{(m)}_{j|i}$, as well as the $t_{j|i}$'s in Eq.~(\ref{tij}) are all between zero and one. To moreover preserve normality of the probability distribution, we moreover require that
\begin{eqnarray} \label{26c}
\sum_{j=1}^N t_{j|i} &=& 1 \, ,
\end{eqnarray}
with the matrix elements $ t_{j|i}$ defined as in Eq.~(\ref{tij}). Hence, the transition matrix $T$ in Eq.~(\ref{TTT}) is an $N \times N$ matrix. Given the $N$ constraints in Eq.~(\ref{26c}), we therefore have $N^2-N=N(N-1)$ free parameters for the matrix $T$. Once $T$ is fixed, the sub-transition matrix $T_A$ can assume $N^2$ positive free parameters but these are bounded from above by the matrix elements of $T$. Once we know $T_A$, we also know $T_B$. Hence, in total, the characterisation of a Mealy HMM requires $N(N-1)+N^2 = (2N-1)N$ independent parameters.

HMMs are non-deterministic (non-unifilar), since the same transition path can result in different stochastic outputs which increases their complexity  \cite{marzen2015informational}. For example, the stationary state is again the distribution vector ${\boldsymbol p}_{\rm ss} $ which is an eigenvector of the transition matrix $T$ such that  $T {\boldsymbol p}_{\rm ss} = {\boldsymbol p}_{\rm ss}$, as stated in Eq.~(\ref{Tpi}). For HMMs, ${\boldsymbol p}_{\rm ss} = (p_{1}, p_{2}, \ldots , p_{N})^{\rm T}$ is an $N$-dimensional column vector. In case of ergodicity, the state vector ${\boldsymbol p}_{\rm ss} $ can be found for example numerically by applying $T$ repeatedly to an initial state until the state of the machine remains the same. But there is no guarantee that a HMM with $N$ internal states has only a single stationary state. For example, an HMM could consists of two independent HMM's with $N_1$ and $N_2 = N - N_1$ internal states. In our analysis of HMMs in the Section \ref{sec4}, we only consider the stationary state ${\boldsymbol p}_{\rm ss}$ which is first calculated in simulations of these machines.

Since the transitions of a HMM are governed by two sub-transitions matrices, namely $T_A$ and $T_B$,  the probability $P(i_1i_2 \ldots i_m)$ for generating the output sequence $i_1 i_2 \ldots i_m$ now reads
 \begin{eqnarray} \label{Pis2}
P(i_1i_2 \ldots i_m) &=& \eta \, T_{i_m} \ldots T_{i_2} T_{i_1} \, {\boldsymbol p}_{\rm ss}
 \end{eqnarray}
where $\eta = (1,1,\ldots,1)^{\rm T}$ is an $N$-dimensional row vector. More concretely, the probability $P_m (A B \ldots BA)$ in Eq.~(\ref{24}) becomes
\begin{eqnarray} \label{24x}
P_m (A B \ldots BA) &=& \eta \, T_A T_B^m T_A \, {\boldsymbol p}_{\rm ss} \, .
\end{eqnarray}
If we ignore the $m$ output symbols between the first and the last $A$, this probability changes into
\begin{eqnarray} \label{24xx}
P_m (A* \ldots *A) &=& \eta \, T_A T^m T_A \, {\boldsymbol p}_{\rm ss} \, .
\end{eqnarray}
All three probabilities differ significantly from the probabilities in Eqs.~(\ref{23})-(\ref{26}). As we shall see in the next section, HMMs are more likely to produce more correlated output sequences than MMs because of their hidden memory.
 
\subsection{Hidden Quantum Markov Models} \label{sec3.1}

Next let us have a closer look at how to parametrise HQMMs. Their generalised measurements can be realised by allowing the qubit, which encodes the hidden state of the machine, to interact with an auxiliary quantum system, i.e.~an environment, followed by projective measurements on a coarse-grained time scale $\Delta t$. In every time step, some hidden information can leak into the environment. Markovianity requires that the ancilla, i.e.~the environment, is reset to the same initial state after each measurement. Otherwise, the dynamics of the HQMM would not depend only on the current state of its memory qubit. Fig.~\ref{fig2}(c) illustrates the stochastic dynamics of the qubit and the random measurement outcomes that might be produced in a single run of such a machine.

\subsubsection{Parametrisation of Kraus operators} \label{IIIC1}

As in Refs.~\cite{monras2010hidden,clark2014hidden}, we model HQMM in the following using the language of open quantum systems and introduce Kraus operators $K_A$ and $K_B$ which we associate with the output symbols $A$ and $B$ of the machine. To parametrise these Kraus operators, we write them in the following as 
\begin{eqnarray} \label{31}
K_m=\begin{pmatrix}
k^{(m)}_{00} & k^{(m)}_{01} \\
k^{(m)}_{10}& k^{(m)}_{11}
\end{pmatrix} \, .
\end{eqnarray}
The eight complex matrix elements of $K_A$ and $K_B$ can be represented by 16 real parameters. However, as pointed out in Eq.~(\ref{constraint}), $K_A$ and $K_B$ must obey a matrix equation which implies that 
\begin{eqnarray} \label{32}
\sum_{m=A,B} |k_{00}^{(m)}|^2+|k_{10}^{(m)}|^2 &=& 1 \, , \notag \\
\sum_{m=A,B} |k_{01}^{(m)}|^2+|k_{11}^{(m)}|^2 &=& 1 \, , \notag \\
\sum_{m=A,B} k_{00}^{(m)} k_{01}^{(m)*} + k_{10}^{(m)} k_{11}^{(m)*} &=& 0 \, .
\end{eqnarray}
These three equations impose four (real) constraints on the above mentioned 16 real parameters, thereby reducing the total number of free (real) parameters needed to fully characterise HQMMs with one memory qubit to 12. This means, a one-qubit HQMM has more free parameters than a  HMM with one or two internal states but less free parameters than a HMM with more than two internal states. Nevertheless, as we shall see in the next section, it is a more powerful stochastic generator.

\subsubsection{Stationary states}

If a track record of all measurement outcomes is kept, a HQMM which has initially been prepared in a pure state $|\psi(0) \rangle$, can always be described by a pure state $|\psi(n) \rangle$. However, as mentioned already in the previous section, if this is not the case and measurement outcomes are ignored, the HQMM must be described by a density matrix $\rho(n)$ instead. How this density matrix evolves from one time step to the next is shown in Eq.~(\ref{qsteps}). Its stationary state is therefore the density matrix $\rho_{\rm ss}$ with
\begin{eqnarray} \label{SS}
{\cal K} ( \rho_{\rm ss}) &=& \rho_{\rm ss} 
\end{eqnarray}
with the superoperator ${\cal K}$ defined in Eq.~(\ref{defK}). Since the HQMM is a two-level system, its stationary state density matrix $\rho_{\rm ss}$ can be written as 
\begin{eqnarray} \label{SS2}
\rho_{\rm ss} &=& \begin{pmatrix}
\rho_{00} & \rho_{01}\\
\rho_{10} & \rho_{11} \end{pmatrix}
\end{eqnarray}
with two real matrix elements, $\rho_{00}$ and $\rho_{11}$, and two complex matrix elements, $\rho_{01}$ and $\rho_{10}$, and with
\begin{eqnarray} \label{35}
\rho_{00} + \rho_{01} &=& 1 \, , \notag \\
\rho_{01} &=& \rho_{10}^* \, . 
\end{eqnarray}
In principle, using Eqs.~(\ref{31})-(\ref{35}), the stationary state density matrix $\rho_{\rm ss}$ of HQMMs could be calculated analytically but the resulting equations do not provide much insight, since they still contain 12 free parameters. In the following section, we therefore solve the above equations only numerically.

\subsubsection{Word probabilities}

As before, we now have a closer look at word probabilities. For example, the probability $P(i_1i_2 \ldots i_m)$ in Eqs.~(\ref{23}) and (\ref{Pis2}) now equals
\begin{eqnarray}
P(i_1i_2 \ldots i_m) &=& \text{Tr} \left( K_{i_m} \ldots K_{i_2} K_{i_1} \, \rho_{\rm ss}  \, K_{i_1}^\dagger K_{i_2}^\dagger \ldots K_{i_m}^\dagger \right) \notag \\
 \end{eqnarray}
with the Kraus operators $K_m$ given in Eq.~(\ref{31}). Moreover, the probabilities $P_m (AB \ldots BA)$ and $P_m (A* \ldots *A)$ are now given by
\begin{eqnarray} 
P_m (AB \ldots BA) &=& \text{Tr} \left( K_A K_B^m K_A \, \rho_{\rm ss} \, K_A^\dagger K_B^{m \, \dagger} K_A^\dagger \right) \, , ~~~ \notag \\ 
P_m(A* \ldots *A) &=& \text{Tr} \left( K_A \, {\cal K}^m \left( K_A \rho_{\rm ss} K_A^\dagger \right)  K_A^\dagger \right) 
\end{eqnarray}
which have many similarities with Eqs.~(\ref{24x}) and (\ref{24xx}). 

\section{A comparison of the complexity of MMs, HMMs and HQMMs} \label{sec4}

In this final section, we use the parametrisation of MMs, HMMs and HQMMs with two output symbols which we introduced in the previous section to study and compare their word probabilities. As we shall see below, there is not much difference between the complexity of MMs and HMMs. Moreover we find that increasing the number of internal states of HMMs does not significantly change the correlation range of their output symbols. The possible correlations between output symbols seem to disappear relatively quickly, after only a few time steps. For simplicity, we only consider a subset of all possible HQMMs with only 3 instead of 12 free parameters. Nevertheless, we find that the HQMMs are the most complex of all three machines. 

\subsection{Simulating HMMs with more than two internal states} \label{sec4.1}

\begin{figure*}[t]
\includegraphics[width=0.9 \textwidth]{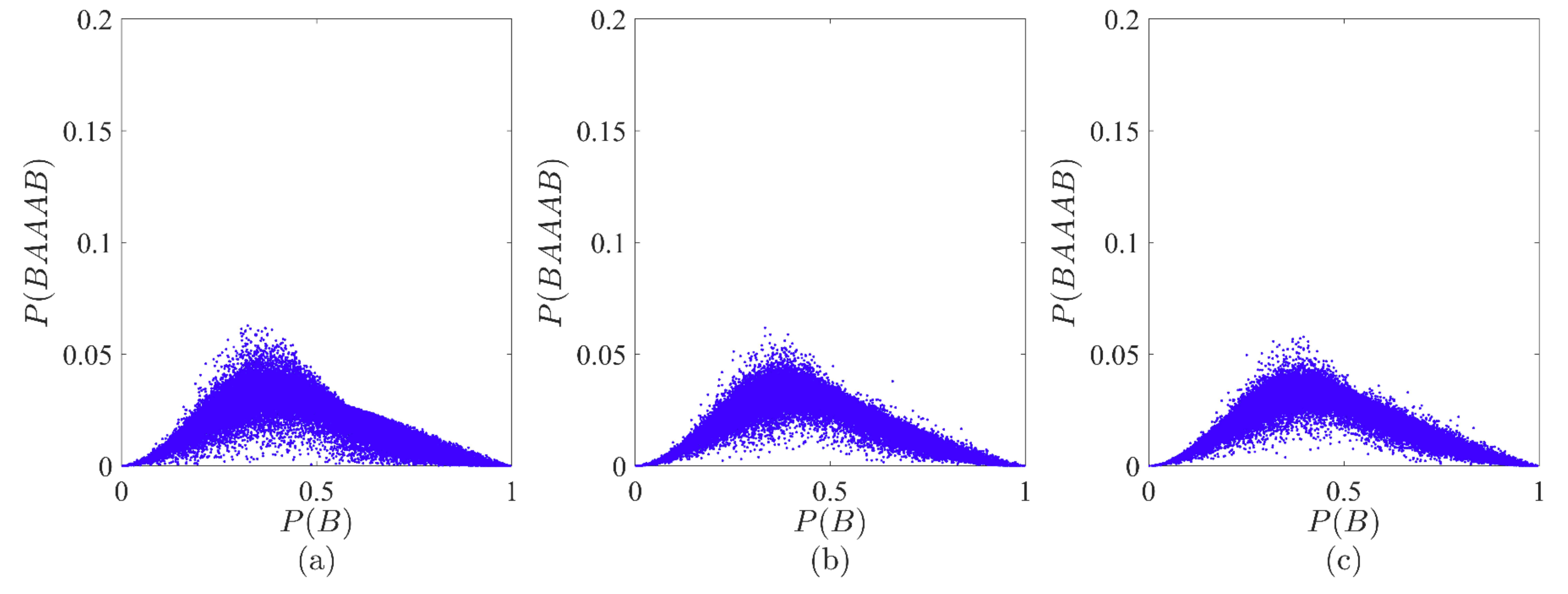}
\caption{The blue dots show the probability $P(BAAAB)$ as function of $P(B)$ for $10^5$ randomly generated stochastic generators which satisfy the parametrisation and conditions imposed on HMMs. The three figures show HMMs with (a) two, (b) three, and (c) four internal states, respectively. For each randomly generated machine, we only calculated  the probabilities $P(BAAAB)$ as function of $P(B)$ of observing the sequence ``$BAAAB$" and a ``$B$" once the respective machine reached a stationary state.} \label{fig:HMM_states}
\end{figure*}

First, let us have a closer look at the effect of increasing the number of internal states $i$ of HMMs on their measurement correlations, while keeping the number of outcomes $A$ and $B$ the same. To do so, we studied the output sequences of HMMs with $2$, $3$ and $4$ internal states. The simulation was carried out as follows: first, we determine a random transition matrix $T$ as well as random sub-transition matrices $T_A$ and $T_B$ so that the elements of each matrix fulfil the constraints imposed by the model. Then, we determine the corresponding stationary state and evaluate the probabilities $P(B)$ and $P(BAAAB)$. Subsequently, these probabilities are calculated for $10^5$ randomly generated machines and used as the coordinates for the blue dots in Fig.~\ref{fig:HMM_states}. 

As shown in Fig.~\ref{fig:HMM_states}, the differences between the three plots are almost negligible. The reason for this might be that we consider machines with only two output symbols. Having only two internal states sees to be sufficient to generate all the correlations that the corresponding HMMs can produce.  Expanding the HMMs to include more than two internal states does not improve the performance of the model. If anything, the space that the blue dots occupy in Figs.~\ref{fig:HMM_states}(b) and (c), respectively, seems to be slightly lower than the space they occupy in Fig.~\ref{fig:HMM_states}(a), due to the increasingly large space needed to sample. Increasing the number of internal states seems to make the generation of an ``optimal" machine less likely. From this we conclude that there is no benefit in considering HMMs with more than two internal states in this paper.

\subsection{An interesting subset of HQMMs}

As we have seen in Section \ref{sec3.1}, a complete parametrisation of 1-qubit HQMMs requires 12 real parameters. However, to conclude that HQMMs are more complex than HMMs, we only need to find one machine that cannot be modelled classically. Keeping this in mind, we consider in the following HQMMs with Kraus operators $K_A$ and $K_B$ which can be written as 
\begin{eqnarray} \label{KAB}
K_A &=& \begin{pmatrix}
\cos \varphi & -a \sin \varphi\\
\sin \varphi & a \cos \varphi
\end{pmatrix} \, , ~~ \notag \\
K_B &=& \begin{pmatrix}
0 & \sqrt{1-a^2} \, \sin \vartheta \\
0 & \sqrt{1-a^2} \, \cos \vartheta 
\end{pmatrix} \, . 
\end{eqnarray}  
Here $a$, $\varphi$ and $\vartheta$ are real parameters with 
\begin{eqnarray} \label{pars2}
a \in (0,1) \, , ~~ \varphi \in (0,2 \pi) \,  , ~~ \vartheta \in (0, 2 \pi) \, .
\end{eqnarray}  
It is relatively straightforward to check that the above operators are indeed valid Kraus operators. Instead of 12, we now only have to deal with three free parameters. The stationary states of the above HQMMs can calculated using Eqs.~(\ref{31}) and (\ref{32}).

\begin{figure*}[t]
\includegraphics[width=0.9 \textwidth]{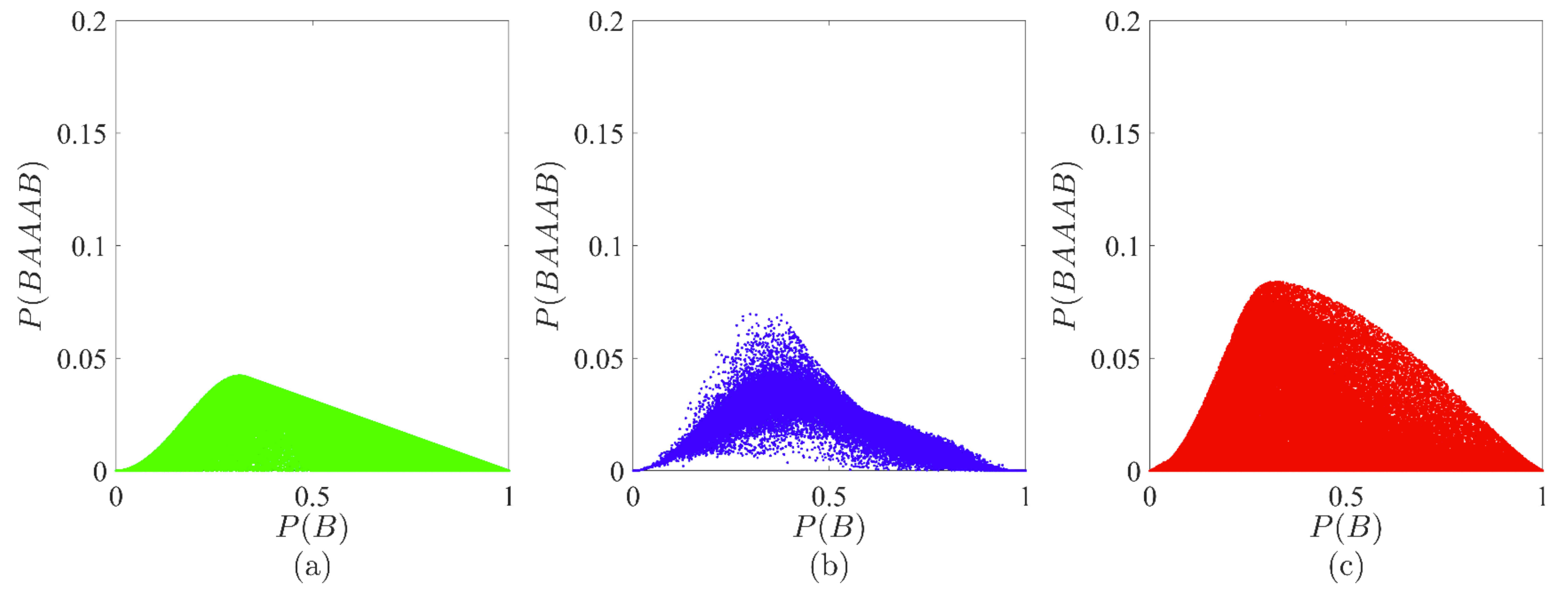}
\caption{The probability $P(BAAAB)$ of observing the sequence ``$BAAAB$" as function of $P(B)$ for (a) a Markov chain (green), (b) a Hidden Markov model (blue) and (c) a Hidden Quantum Markov model(red). Each point corresponds to a randomly generated machine. As one might expect, the comparison of the figures shows in general higher correlations between the measurement outcomes for HQMM than for their classical counterparts, HMMs and MMs, since the red dots cover a much bigger area. For each figure, $10^5$ randomly generated machines were taken into account.}\label{fig:mc_hmm_hqmm}
\end{figure*}

To show that the above-described generalised measurement is of physical relevance, let us have a closer look at how it could be realised experimentally. As an example, we assume that the memory qubit is a single atom with ground state $|1 \rangle$ and an excited state $|2 \rangle$.  Suppose the initial state of the atom equals 
\begin{eqnarray}
|\psi (0) \rangle &=& c_1 \, |1 \rangle + c_2 \, |2 \rangle  
\end{eqnarray}
and $\Gamma $ is the spontaneous decay rate of its excited state. Under the condition of no photon emission, the unnormalised state of the atom equals \cite{Heger}
\begin{eqnarray}
|\psi (\Delta t) \rangle &=& c_1 \, |1 \rangle + {\rm e}^{-\Gamma \, \Delta t/2} \, c_2 \, |2 \rangle 
\end{eqnarray} 
after some time $\Delta t$, since not seeing a photon reveals information about the state of the atom and decreases the probability that the atom is in its excited state. Suppose, we rotate the state of the atom at time $\Delta t$ by applying the unitary operator 
\begin{eqnarray}
U_A &=& \begin{pmatrix}
\cos \varphi & - \sin \varphi\\
\sin \varphi & \cos \varphi
\end{pmatrix} \, .
\end{eqnarray} 
This can be done, for example, using a short strong laser pulse. It is relatively straightforward to check that the overall effect is the implementation of the Kraus operator $K_A$, if the parameter $a$ in Eq.~(\ref{KAB}) equals $\exp(-\Gamma \, \Delta t/2)$. This applies, since 
\begin{eqnarray}
K_A \, |\psi (0) \rangle &=& U_A \, |\psi (\Delta t) \rangle 
\end{eqnarray}
in this case. In other words, not seeing a photon in a time interval $(0,\Delta t)$ followed by the application of $U_A$ is a way of realising $K_A$.

However, if the atom emits a photon within the time interval $(0,\Delta t)$, its unnormalised state equals \cite{Heger}
\begin{eqnarray}
|\psi (\Delta t) \rangle &=& \sqrt{1-{\rm e}^{-\Gamma \, \Delta t}} \, c_1 \, |1 \rangle \, .
\end{eqnarray} 
Now suppose, in this case, with apply the unitary operator
\begin{eqnarray}
U_B &=& \begin{pmatrix}
\cos \vartheta & \sin \vartheta\\
- \sin \vartheta & \cos \vartheta
\end{pmatrix} 
\end{eqnarray} 
at time $\Delta t$. Then seeing a photon changes the state of the atom, up to a normalisation factor, into $K_B  \, |\psi (0) \rangle$, since 
\begin{eqnarray}
K_B \, |\psi (0) \rangle &=& U_B \, |\psi (\Delta t) \rangle 
\end{eqnarray}
in this case. This means, seeing a photon now constitutes a $B$ measurement. One can easily check that the probabilities of getting the measurement outcomes $A$ and $B$ are as one would expect for a generalised measurement with Kraus operators $K_A$ and $K_B$. In the next subsection, we have a closer look at the output sequences that can be obtained by applying the above-described generalised measurement repeatedly to a memory qubit.

\subsection{A comparison of the performance of the machines} \label{sec3.3}
 
This final subsection outlines a comparative analysis of the stochastic properties of the output sequences of the above-introduced MMs, HMMs and HQMMs with output symbols $A$ and $B$. In each case, we generated a large number of random machines and determined their stationary-state word probabilities $P(BAAAB)$ and $P(B)$ as previously described, for example, in Section \ref{sec4.1}. For each machine, when then place a single dot in the corresponding figure. The selected sequence for the simulation is ``$BAAAB$", representing the probability of observing the event $B$ subsequent to three consecutive occurrences of $AAA$, given that the initial outcome is $B$. In other words, this sequence provides insight into the correlation between two $B$ outcomes separated by multiple observations of $A$. Fig.~\ref{fig:mc_hmm_hqmm} illustrates that the HQMMs are much more likely to generate word sequences ``$BAAAB$" for a given probability $P(B)$. This means, they exhibit a superior performance due to their ability to occupy a larger state space compared to their classical counterparts, thereby maintaining some information about the history of the machines despite having the Markov property. In summary, despite utilising only a single qubit as a memory, HQMMs have the capacity to exhibit temporal correlations which cannot be reproduced by linear hidden variable models. 

\section{Conclusions} \label{sec5}

This paper compares classical Markov chain-based models, specifically Markov Models (MMs) and Hidden Markov Models (HMMs) with their quantum counterparts, so-called Hidden Quantum Markov Models (HQMMs). We reviewed the definitions of the three models with a specific emphasis on their parametrization. When increasing the internal states of HMMs with two outputs $A$ and $B$, we observed that this does not improve the correlations among output sequences. From this we concluded that HMMs with two outputs do not benefit from having more than two internal states. We then examined the probability of observing a specific output sequence ``$BAAAB$" for a large number of randomly generated machines. Our simulations of the corresponding word probabilities $P(BAAAB)$ shows that HQMMs can exhibit superior performance. For certain examples of HQMMs, the probabilities $P(BAAAB)$ are larger than the maximum probability $P(BAAAB)$ that can be realised with linear classical models. This means, quantum physics cannot be captured by classical linear hidden variable theories even in the absence of entanglement. Our work emphasizes the quantumness of temporal correlations in the output sequences produced by generalised measurements and recommends them as an alternative resource (which is different from entanglement and usually much easier to produce) for quantum technology applications. \\[0.5cm]
{\bf Acknowledgement.} LAC acknowledges support from the Foundation for Polish Science within the “Quantum Optical Technologies” project carried out within the International Research Agendas programme co-financed by the European Union under the European Regional Development Fund. KAR acknowledges the support from the Ministry of Higher Education, Research and Innovation in the Sultanate of Oman through The National Postgraduate Scholarship Programme.

\end{document}